\documentclass[prd,aps,showpacs,floats,letterpaper,floatfix,
groupedaddress,superscriptaddress
,nofootinbib
,eqsecnum,
twocolumn
]{revtex4}
\usepackage{float}
\usepackage{placeins}
\restylefloat{table}
\usepackage{comment}
\usepackage{amsmath}
\usepackage{amsfonts}
\usepackage{bm}
\usepackage{graphicx}
\usepackage{wrapfig}
\usepackage{appendix}
\usepackage{etoolbox}
\BeforeBeginEnvironment{appendices}{\clearpage}
  \newcommand{\mnras}{Mon. Not. R. Astron. Soc.}

   \newcommand{\lrr}{Living Rev. Relativ.}
   
   \newcommand{\aj}{Astron. J.}
 \usepackage[dvipsnames]{xcolor}

\def\v1v2{{\bf v}_1 \cdot {\bf v}_2}

\usepackage[super]{nth}
\usepackage[nameinlink]{cleveref}

\allowdisplaybreaks

\begin{document}

\title{Residual eccentricity of inspiralling orbits at the gravitational-wave detection threshold: Accurate estimates using post-Newtonian theory}

\author{
Alexandria Tucker} \email{a.tucker@ufl.edu}
\affiliation{Department of Physics, University of Florida, Gainesville, Florida 32611, USA}

\author{
Clifford M. Will} \email{cmw@phys.ufl.edu}
\affiliation{Department of Physics, University of Florida, Gainesville, Florida 32611, USA}
\affiliation{GReCO, Institut d'Astrophysique de Paris, CNRS, Sorbonne Universit\'e \\ 
98 bis Bd. Arago, 75014 Paris, France}

\date{\today}

\begin{abstract}
We use equations of motion containing gravitational radiation-reaction terms through 4.5 post-Newtonian order to calculate the late-time eccentricities of inspiraling binary systems of non-spinning compact bodies as they cross the detection threshold of ground-based gravitational-wave interferometers.  The initial eccentricities can be as large as $0.999$.   We find that the final eccentricities are systematically smaller than those predicted by the leading quadrupole approximation, by as much as 30 percent for a 300 solar mass binary crossing the LIGO/Virgo detection threshold at $10$ Hz, or eight percent smaller for a 60 solar mass binary.  We find an analytic formula for the late-time eccentricity that accurately accounts for the higher-order post-Newtonian effects, generalizing a formula derived by Peters and Mathews in the 1960s.   We also find that the final eccentricities are independent  of the ratio of the masses of the two compact bodies to better than two percent.

\end{abstract}

\maketitle

\section{Introduction and Summary}
\label{sec:intro}

A notable fact about gravitational radiation from binary systems is that it takes angular momentum away from the system as effectively as it takes away energy.  A consequence is that as the binary system shrinks, or inspirals, it circularizes, and this effect is so pronounced that the orbits will generally become extremely circular long before they coalesce.  Consider for example, the first detected inspiraling binary system, the Hulse-Taylor binary pulsar B1913+16.  Its orbit has a rather large orbital eccentricity of $0.617$.  But by the time its gravitational wave signals cross the LIGO/Virgo detection threshold of around $10$ Hz, about 390 million years from now, its eccentricity will be only about $5 \times 10^{-6}$.

This fact led to the expectation, borne out by experience, that the initial gravitational wave signals detected by the LIGO/Virgo network would be from compact binary sources whose orbits were essentially perfectly circular, apart from their monotonic inspiral (see, however \cite{gayathri2020gw190521,gamba2021gw190521} for preliminary evidence of a highly eccentric merger in GW190521).

But this assumes systems that undergo isolated evolution from wide, non-relativistic orbits over long periods of time.  However, there is reason to expect that some detectable gravitational wave signals will not come from such pristine sources.  A system could be formed in a tight but highly eccentric orbit by three-body processes that send one body far from the system, leaving a tight binary remnant \cite{1975MNRAS.173..729H}, or by direct capture from an unbound orbit into a bound eccentric orbit by gravitational-wave emission during the close passage \cite{1987ApJ...321..199Q}.  In a hierarchical three-body system consisting of a compact binary in the presence of a distant third body, the famous Kozai-Lidov oscillations \cite{1962AJ.....67..591K,1962P&SS....9..719L} could drive the binary's eccentricity to large values, even as gravitational-wave emission causes the orbit to shrink and attempts to circularize it \cite{2003ApJ...598..419W,2013ApJ...773..187N,2016ApJ...828...77V,2017ApJ...836...39S,2018ApJ...856..140H,2018ApJ...853...93R,2018ApJ...864..134R}.

The idea that compact binary systems could enter the LIGO/Virgo band with non-trivial orbital eccentricity has two implications:

\begin{enumerate}
\item
The theoretical template gravitational waveforms used in the initial detections and analyses were based on quasicircular models for the orbits.  Such templates may not be as effective in detecting signals from eccentric inspirals, and may introduce biases in estimating the parameters of the sources.  Accordingly, the construction of eccentric templates is important, and considerable effort in this direction is ongoing \cite{2009PhRvD..80h4001Y,2009PhRvD..80l4018A,2010PhRvD..82l4064T,2011AnP...523..813T,2012PhRvD..86j4027M,2013PhRvD..87l7501H,2014PhRvD..90h4016H,2015PhRvD..91f3004C,2015PhRvD..92d4034S,2016PhRvD..93d3007T,2016PhRvD..93f4031T,2017PhRvD..95b4038H,2017PhRvD..96d4028C,2017PhRvD..96j4048H,2018PhRvD..97b4031H,2018PhRvD..98j4043K,2018CQGra..35w5006M,2019PhRvD..99l4008T,2019CQGra..36r5003M,2019PhRvD.100d4018B,2019PhRvD.100h4043E,2019PhRvD.100f4006T,2020PhRvD.101d4049L}.

\item
Conversely, the detection and measurement of eccentric inspiral events could serve to confirm or distinguish among various proposed astrophysical formation channels for  these inspiralling compact binaries (see \cite{favata2021constraining} for discussion and references).  This only works, however, if the residual eccentricities are above a reasonable detection threshold.

\end{enumerate}  

Our goal in this paper is to provide an accurate map from the initial parameters of an arbitrarily eccentric binary orbit to the orbital eccentricity when the gravitational wave frequency reaches a detection threshold for a given detector.  We imagine a binary consisting of two non-spinning ``point'' bodies of arbitrary mass whose evolution is dominated by two-body relativistic effects, i.e. in which other perturbations, such as Kozai-Lidov effects, can be ignored.  We will be concerned only with the long-term evolution of such systems leading up to their entering the sensitive band of the detector, not with the subsequent evolution leading to merger and ringdown.  We ignore finite-size effects, such as tidal interactions, as these are relevant mainly for the late-time, highly relativistic regime.  We also ignore spin effects.

Such a map already exists.  It is based on the classic 1963-1964 papers by Philip Peters and Jon Mathews \cite{1963PhRv..131..435P,1964PhRv..136.1224P}, who computed the energy and angular momentum flux due to gravitational waves at the quadrupole order of approximation, leading to the map (see Eq.\ (5.11) of \cite{1964PhRv..136.1224P})
\begin{equation}
p = p_i \frac{g(e)}{g(e_i)}  \,, 
\label{eq:PM}
\end{equation}
where 
\begin{equation}
g(e) = e^{12/19} (304+121 e^2)^{870/2299} \,,
\end{equation}
and
where $p_i$ and $e_i$ are the initial semilatus rectum and eccentricity of the orbit and $p$ and $e$ are the values at a later time.  Recall that $p$ is related to the Newtonian orbital angular momentum, and that $p=a(1-e^2)$ where $a$ is the semimajor axis, related to the Newtonian energy.   

We improve this map by incorporating post-Newtonian (PN) corrections.  We use equations of motion for the binary system that include conservative terms through third post-Newtonian (3PN) order, and radiation-reaction terms through 4.5PN order, including the leading 4PN ``tail'' terms.  We then obtain the long-term, orbit-averaged evolution equations for $p$ and $e$.   On integrating these equations numerically, we find that the late-time values of eccentricity are independent (to better than two percent) of the value of the symmetric mass ratio 
$\eta \equiv m_1 m_2/m^2$, where $m=m_1+m_2$.  Thus, whereas the factor $\eta$ controls the {\em rate} of emission of energy and angular momentum and thus the lifetime of the system (with $T \propto 1/\eta$), it has essentially no effect on the relation between eccentricity and semilatus rectum.   This relation is then as valid for equal-mass binary inspirals as for extreme mass-ratio inspirals (EMRIs).

\begin{figure}[t]
\begin{center}

\includegraphics[width=3.4in]{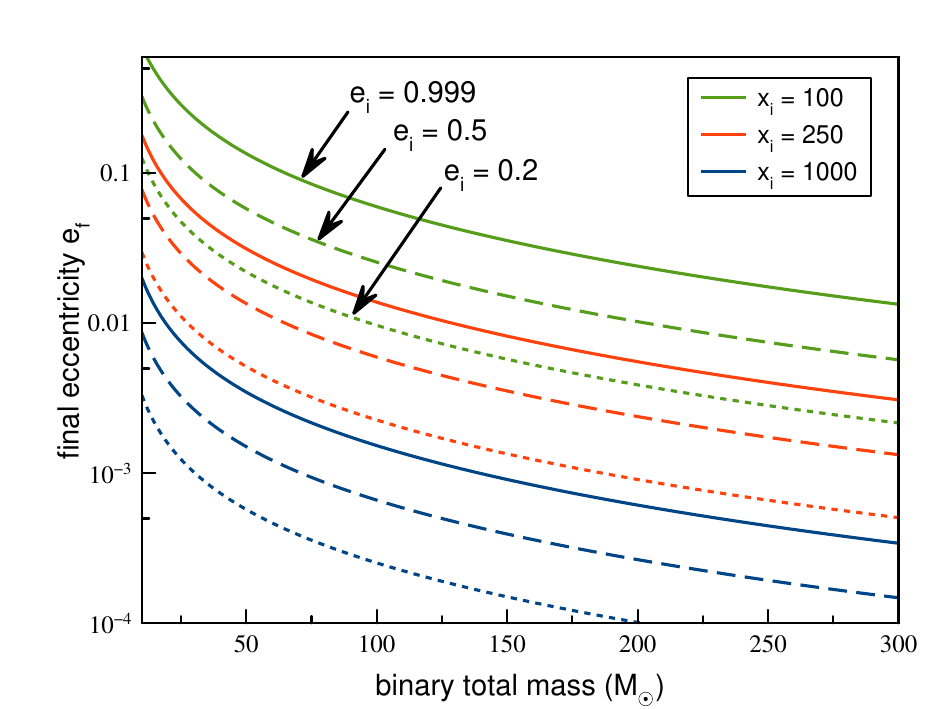}

\caption{\label{fig:efinmass} Final eccentricity vs. total mass of the binary system, for LIGO/Virgo sources, assuming a 10 Hz threshold detection frequency.  Solid, dashed and dotted lines correspond to initial eccentricities of 0.999, 0.5 and 0.2, respectively.  Green, red and blue curves correspond to $p_i = 100$, $250$ and $1000$ $Gm/c^2$, respectively.  For a third generation detector with a threshold detection frequency $f_{\rm 3G}$, multiply the masses by $10/f_{\rm 3G}$.  }
\end{center}
\end{figure}

Using the approximate relation between gravitational wave frequency $f$ and semilatus rectum $p$ in the limit of small eccentricity,  $p \approx (Gm)^{1/3}/(\pi f)^{2/3}$, we plot the expected eccentricity as a function of the mass of the binary, assuming a threshold detection frequency relevant to advanced LIGO/Virgo of $10$ Hz.  The results are shown in Fig.\ \ref{fig:efinmass}. 
 We choose three sets of initial values for $x_i= c^2 p_i/Gm$: 100, 250, and 1000, and three initial eccentricities, 0.999, 0.5 and 0.2.   For larger initial values of $p$, the residual eccentricities are smaller, reflecting the increased time for radiation reaction to circularize the orbit.  The larger the mass of the binary, the smaller the residual eccentricity, because higher mass systems cross the $10$ Hz detection threshold at smaller values of separation (smaller values of $p$), thus after additional circularization has occurred.   Consider, for example, a $20 \, M_\odot$, equal-mass binary system with the initial values $p_i = 250 (Gm/c^2)$ and $e_i = 0.999$, corresponding to a pericenter separation of $3500$ km and an apocenter separation of almost 7.5 million km.  By the time the binary enters the LIGO/Virgo band around 13 days later, its eccentricity will have decreased to $0.086$, illustrating the strong circularization property of gravitational radiation reaction.
 
Notice that the final eccentricity depends on $x_f \approx (c^3/\pi Gm f_{\rm th} )^{2/3}$, where $f_{\rm th}$ is the threshold detection frequency.  As a result, should that frequency be reduced from $10$ Hz to a lower frequency $f_{\rm 3G}$, say in third-generation detectors such as ET or Cosmic Explorer, then the final eccentricities are still given by Fig.\ \ref{fig:efinmass}, but with the mass values on the horizontal axis multiplied by $10/f_{\rm 3G}$.
 
\begin{figure}[t]
\begin{center}

\includegraphics[width=3.4in]{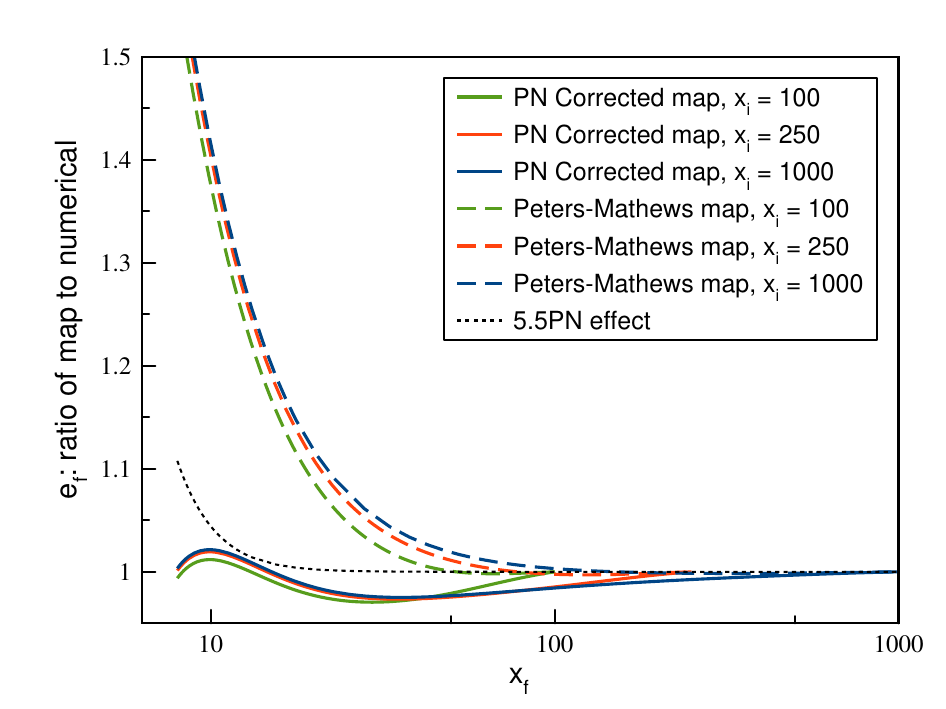}

\caption{\label{fig:EfinalFits} Analytic fits to the final eccentricity.  Solid curves are the values of $e_f$ obtained from Eq.\ (\ref{eq:PNcorrected}), while dashed curves are values obtained from the Peters-Mathews map (\ref{eq:PM}), all normalized to the numerical results.  The black dotted curve shows the (negative) of the effect on the numerical results of the 5.5PN corrections in the equations of motion.    }
\end{center}
\end{figure}

After some experimentation, we have found a simple, analytic, PN-corrected map that fits these numerical results to better than two percent over the relevant range of parameters.  In terms of the dimensionless semilatus rectum $x = c^2 p/Gm$, it is given by
\begin{align}
x &=x_i \left ( \frac{1+2/x_i}{1+2/x} \right ) \left 
( \frac{1-4/x_i}{1-4/x} \right )^{12/19} \frac{g(e)}{g(e_i)}
\,.
\label{eq:PNcorrected}
\end{align}

Figure \ref{fig:EfinalFits} displays values of $e_f$ from the PN-corrected map (\ref {eq:PNcorrected}) normalized to the numerical results, as a function of $x_f$.  The agreement is better than two percent over the entire range of $x_f$.  Three initial values of $x$: 100, 250 and 1000 are displayed; in all cases shown, the initial eccentricity is 0.999.  For comparison, we show the Peters-Mathews (PM) map, also normalized to the numerical values.  From these results we conclude that {\em the Peters-Mathews formula consistently overestimates the late-time eccentricity}.  
Also plotted in black dots is a curve showing the effect of adding 5.5PN terms to the equations of motion, as a way to illustrate the rough errors in the PN approximation.  We discuss our tests of the  validity of the PN approximation in more detail in Sec.\ \ref{sec:PNconvergence}.  

The remainder of the paper provides the details behind these results.  In Sec.\ \ref{sec:evolution}, we write down the post-Newtonian equations of motion to the order used in our analysis, obtain the Lagrange planetary equations for the evolution of the osculating orbit elements of the binary, and find the long-term evolution equations for the orbit  elements using a ``two timescale'' approach.
In Sec.\ \ref{sec:numerical}, we integrate the evolution equations numerically and explore the dependence of the relationship between eccentricity and semilatus rectum on the symmetric mass ratio $\eta$ and on the post-Newtonian order used.  We then obtain our PN-corrected map, as well as an accurate analytic approximation for the evolution time.  Final remarks are made in Sec.\ \ref{sec:conclusions}.   In the Appendices we display the coefficients that appear in the post-Newtonian equations of motion, and describe the calculation of the tail term and 5.5PN effects on the long-term evolution of the orbit elements.

\section{Evolution of compact binary orbits to high post-Newtonian order} 
\label{sec:evolution}

\subsection{Equations of motion}
\label{sec:eom}

We analyze the orbital evolution of a binary system of compact, non-spinning bodies in the post-Newtonian approximation. We work in harmonic coordinates, the natural basis for post-Newtonian theory, gravitational radiation and radiation reaction (see \cite{PW2014} for a pedagogical treatment of post-Newtonian theory).   Letting $\bm{r}_a$ and $m_a$ respectively denote the position and mass of body $a\in \{1,2\}$, we define the relative position vector $\bm{r}\equiv\bm{r}_1-\bm{r}_2$ and unit vector $\bm{n}=\bm{r}/r$ pointing from body 2 to body 1, the total mass $m\equiv m_1+m_2$, and the symmetric mass ratio $\eta=m_1m_2/m^2$. 
The relative velocity, acceleration and angular momentum vectors are $\bm{v}=d \bm{r}/dt$, $\bm{a}=d \bm{v}/dt$, and $\bm{h} = \bm{r} \times \bm{v}$.
The equations of motion in terms of relative coordinates take the general form 
\begin{align}
	\bm{a}=& -\frac{Gm}{r^2}\,\bm{n} +			
	   \frac{G m}{r^2}\left(\mathcal{A}_{\mathrm{c}} \,\bm{n} + \frac{1}{\dot{r}} \mathcal{B}_{\mathrm{c}} \,\bm{v} \right )
\nonumber \\
&
 +\frac{8}{5}\eta  \frac{G m}{r^2} \frac{Gm}{rc^3} \left( \dot{r}  \mathcal{A}_{\mathrm{rr}}  \, \bm{n}+
       \mathcal{B}_{\mathrm{rr}} \, \bm{v}  \right)  + \bm{a}_{\rm Tail} \,.
	\label{eq:EOMGen}
\end{align}
Here, $\mathcal{A}_{\mathrm{c}} = \mathcal{A}_{\mathrm{c}}^{(1)} +\mathcal{A}_{\mathrm{c}}^{(2)} +\mathcal{A}_{\mathrm{c}}^{(3)}$  and $\mathcal{B}_{\mathrm{c}} = \mathcal{B}_{\mathrm{c}}^{(1)} +\mathcal{B}_{\mathrm{c}}^{(2)} +\mathcal{B}_{\mathrm{c}}^{(3)}$ denote the conservative contributions at 1PN, 2PN, and 3PN orders, respectively, and
$\mathcal{A}_{\mathrm{rr}} = \mathcal{A}_{\mathrm{rr}}^{(1)} +\mathcal{A}_{\mathrm{rr}}^{(2)} +\mathcal{A}_{\mathrm{rr}}^{(3)}$  and $\mathcal{B}_{\mathrm{rr}} = \mathcal{B}_{\mathrm{rr}}^{(1)} +\mathcal{B}_{\mathrm{rr}}^{(2)} +\mathcal{B}_{\mathrm{rr}}^{(3)}$
 denote radiation reaction terms at 2.5PN, 3.5PN, and 4.5PN orders respectively.  To the orders of interest, the $\cal{A}$ and $\cal{B}$ coefficients take the general forms
\begin{align}
\mathcal{A}_{\rm c}^{(N)} & 
= \sum_{l,m,n} a^{(N)}_{lmn} \frac{ \delta_{l+m+n,N}}{c^{2N}} \left(\frac{Gm}{r}\right)^l \left(\dot{r}^2\right)^m \left(v^2\right)^n \,,
\nonumber \\
\mathcal{B}_{\rm c}^{(N)} & 
= \sum_{l,m,n} b^{(N)}_{lmn} \frac{ \delta_{l+m+n,N}}{c^{2N}} \left(\frac{Gm}{r}\right)^l \left(\dot{r}^2\right)^m \left(v^2\right)^n  \,,
\nonumber \\		
\mathcal{A}_{\rm rr}^{(N)} & 
= \sum_{l,m,n} c^{(N)}_{lmn} \frac{ \delta_{l+m+n,N}}{c^{2N}} \left(\frac{Gm}{r}\right)^l \left(\dot{r}^2\right)^m \left(v^2\right)^n \,, 
\nonumber \\
\mathcal{B}_{\rm rr}^{(N)} & 
= \sum_{l,m,n} d^{(N)}_{lmn} \frac{ \delta_{l+m+n,N}}{c^{2N}} \left(\frac{Gm}{r}\right)^l \left(\dot{r}^2\right)^m \left(v^2\right)^n \,.	
\end{align}
The explicit expressions for the coefficients $\{a,b,c,d\}^{(N)}_{lmn}$ for general mass ratios are given in Appendix \ref{app:coeffs}.

The final term in Eq.\ (\ref{eq:EOMGen}) is the lowest order 4PN tail term, given formally by \cite{1988PhRvD..37.1410B,2000PhRvD..62l4015P}
\begin{equation}
a_{\rm Tail}^j =- \frac{4G^2}{5c^8} m r^k \int_0^\infty \stackrel{(7)\qquad \qquad}{{\cal I}^{\langle jk \rangle}(t -s)} 
\ln (s/2) ds \,,
\label{eq:tailterm}
\end{equation}
where, to the required PN order,
\begin{equation}
{\cal I}^{\langle jk \rangle} \equiv  \eta m \left ( r^j r^k - \frac{1}{3}r^2 \delta^{jk} \right )
\label{eq:quadmom}
\end{equation}
is the trace-free quadrupole moment of the system, the symbol $(7)$ atop $\cal I$ denotes seven time derivatives, and the integral is over the past history of the binary system.  The explicit calculation of the effect of the tail term on the evolution of the orbital elements of the binary will be carried out in Appendix \ref{app:tails}.

\subsection{Osculating orbits and the perturbed Kepler problem}

\begin{figure}[t]
\begin{center}

\includegraphics[width=3.5in]{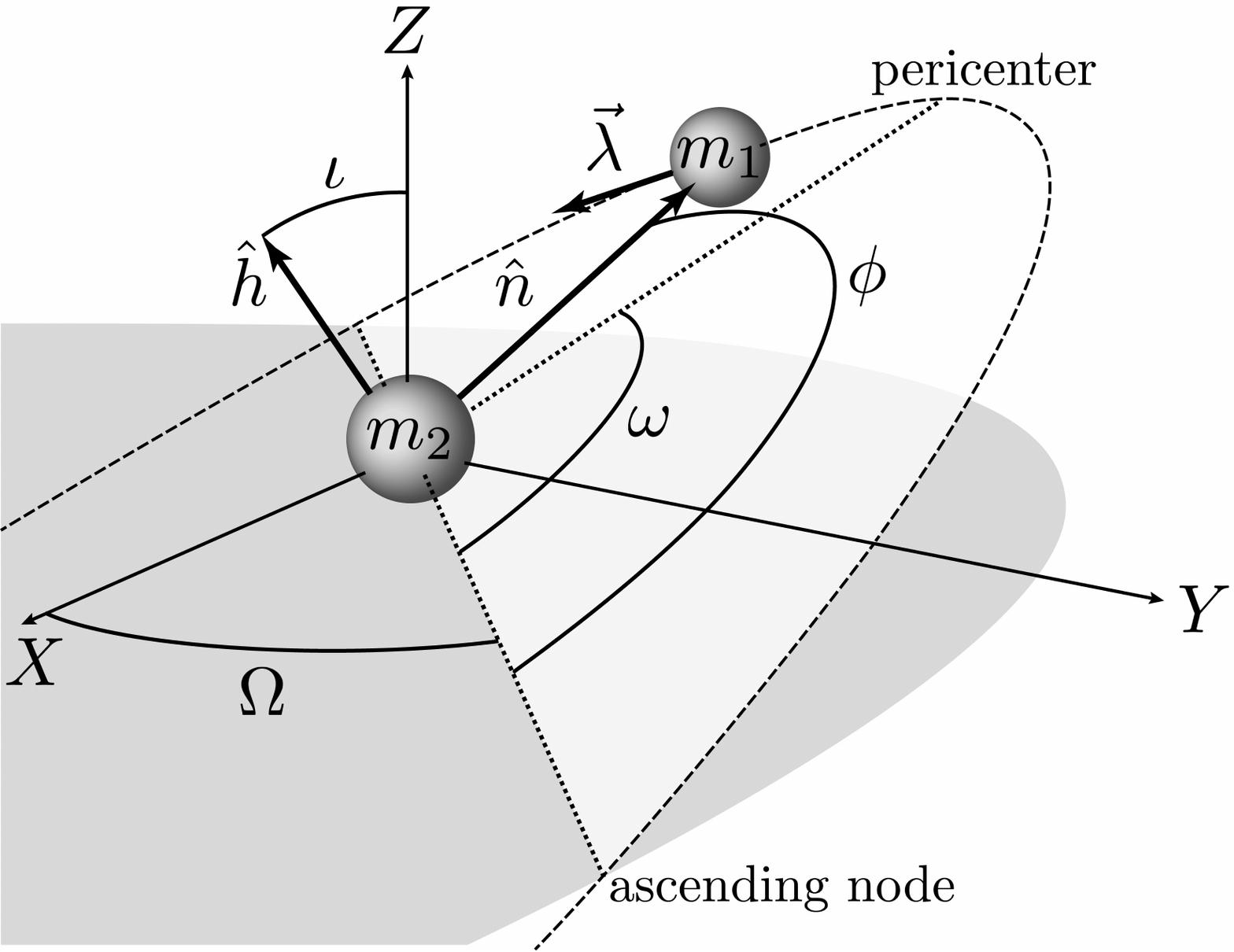}

\caption{\label{fig:orbit} Orbital elements of the effective one-body orbit of a binary system. }
\end{center}
\end{figure}

We employ the ``osculating orbit'' approach to solving the perturbed Kepler problem (see, for example  \cite{PW2014}). In this method, the equations of motion (\ref{eq:EOMGen})  are summarized as 
\begin{equation}
	\bm{a} = -\frac{Gm}{r^2}\bm{n} + \delta \bm{a} \,.
	\end{equation}
The perturbed orbit is defined by six ``osculating'' orbit elements, given by the elements of a pure Keplerian orbit that is momentarily ``tangent'' to the perturbed orbit, i.e.\ that has the same momentary values of $\bm{r}$ and $\bm{v}$.  Because the orbit is perturbed, these osculating elements are no longer constant in time (see Fig.\ \ref{fig:orbit}).  
They are the semilatus rectum $p$, eccentricity $e$, orbital inclination $\iota$, nodal angle $\Omega$, pericenter angle $\omega$ and time of pericenter passage $T$ (this sixth element will not be relevant for our purposes).  They can be defined by the following set of equations
\begin{align}
{\bm r} &\equiv p{\bm n} /(1+e \cos f) \,,
\nonumber \\
{\bm n} &\equiv \left ( \cos \Omega \cos \phi- \cos \iota \sin \Omega \sin \phi \right ) {\bm e}_X 
\nonumber \\
& \quad
 + \left ( \sin \Omega \cos \phi + \cos \iota \cos \Omega \sin \phi \right ){\bm e}_Y
\nonumber \\
& \quad
+ \sin \iota \sin \phi {\bm e}_Z \,,
\nonumber \\
{\bm \lambda} &\equiv \partial {\bm n}/\partial \phi \,, \quad \hat{\bm h}={\bm n} \times {\bm \lambda} \,,
\nonumber \\
{\bm h} &\equiv {\bm r} \times {\bm v} \equiv \sqrt{Gmp} \, \bm{\hat{h}} \,,
\nonumber \\
\dot{r} &\equiv (he/p) \sin f \,,
\label{eq2:keplerorbit}
\end{align}
where  $f \equiv \phi - \omega$ is the  {\em true anomaly}, $\phi$ is the orbital phase measured from the ascending node, and 
 ${\bm e}_A$ are chosen reference basis vectors.   From the given definitions, we see that ${\bm v} = \dot{r} {\bm n} + (h/r) {\bm \lambda}$.  To avoid the well-known problem of the singular behavior of $\omega$ in the circular limit, we will use the alternative orbit elements:
\begin{equation}
	\alpha  \equiv e \cos{\omega}, \quad  \beta \equiv e \sin{\omega} \,,
	\label{eq:alphabeta}
\end{equation}
which are regular when $e\rightarrow 0$.

We then compute the radial $\mathcal{R}\equiv \bm{n} \cdot \delta \bm{a}$, cross-track $\mathcal{S}\equiv \bm{\lambda} \cdot \delta \bm{a} $, and out-of-plane $\mathcal{W}\equiv \hat{\bm{h}} \cdot \delta \bm{a}$ components of the perturbing acceleration.  Because the perturbations lie within the orbital plane,  ${\cal W}  = 0$. 

The ``Lagrange planetary equations'' then describe the evolution of the orbital parameters in response to the perturbing accelerations:
\begin{align}
\frac{dp}{d\phi} &= 2 \frac{r^3}{Gm} \, {\cal S}\,,
\nonumber \\
\frac{d\alpha}{d\phi} &= \frac{r^2}{Gm} \left [ {\cal R}  \sin \phi   +  {\cal S} (\alpha + \cos \phi) \left ( 1+ \frac{r}{p} \right ) 
\right ],
\nonumber \\
\frac{d\beta}{d\phi} &= \frac{r^2}{Gm} \left [ -{\cal R}  \cos \phi   +  {\cal S} (\beta + \sin \phi) \left ( 1+ \frac{r}{p} \right ) 
\right ],
\nonumber \\
\frac{d\iota}{d\phi} &=0\,,
\nonumber \\
\frac{d\Omega}{d\phi} &=0\,,
\label{eq2:Lagrange}
\end{align}
together with the relationship between $\phi$ and time,
\begin{equation}
\frac{d\phi}{dt} = \frac{h}{r^2} \,.
\end{equation}
This simple relation follows from the fact that $\Omega$ and $\iota$ are constant.

\subsection{Two-timescale Analysis}
\label{sec:twoscale}

The Lagrange planetary equations are of the general form
\begin{equation}
\frac{d X_\gamma (\phi)}{d\phi} = \epsilon Q_\gamma (X_\delta(\phi), \phi) \,,
\label{eq2:dXdf}
\end{equation}
where $\gamma, \, \delta$ label the orbit element, and $\epsilon$ is a small parameter that characterizes the perturbation.
We anticipate that the solutions for the $X_\gamma$ will have pieces that vary on a ``short'' orbital timescale, corresponding to the periodic functions of $\phi$, and pieces that vary on a long timescale characterized by a variable $\theta \equiv \epsilon \phi$.   In a two-timescale analysis \cite{1978amms.book.....B,1990PhRvD..42.1123L,2004PhRvD..69j4021M,2008PhRvD..78f4028H,2017PhRvD..95f4003W}, one treats these two times formally as independent variables, and solves the ordinary differential equations as if they were partial differential equations for the two variables.
We  write the derivative with respect to $\phi$ as
\begin{equation}
\frac{d}{d\phi} \equiv \epsilon \frac{\partial}{\partial \theta} + \frac{\partial}{\partial \phi} \,,
\label{eq2:ddphi}
\end{equation}
and make an {\em ansatz} for the solution for $X_\gamma (\theta, \phi)$:
\begin{equation}
X_\gamma (\theta, \phi) \equiv \tilde{X}_\gamma (\theta) + \epsilon Y_\gamma (\tilde{X}_\delta (\theta), \phi) \,.
\label{eq2:ansatz}
\end{equation}
The split is defined by 
\begin{equation}
\tilde{X}_\gamma (\theta) = \langle X_\gamma (\theta, \phi) \rangle \,, \quad \langle Y_\gamma (\tilde{X}_\delta (\theta), \phi) \rangle = 0 \,,
\label{eq2:split}
\end{equation}
where the ``average'' $\langle \dots \rangle$ is defined by 
\begin{equation}
\langle A \rangle \equiv \frac{1}{2\pi} \int_0^{2\pi} A(\theta,\phi) d\phi \,,
\label{eq2:averagedef}
\end{equation}
holding $\theta$ fixed.  
For any function $A(\theta,\phi)$ we define the ``average-free'' part as
\begin{equation}
 {\cal AF}(A) \equiv  A(\theta,\phi) - \langle A \rangle   \,.
 \label{eq2:averagefreedef}
\end{equation}
We now substitute Eqs.\ (\ref{eq2:ddphi}) and (\ref{eq2:ansatz}) into (\ref{eq2:dXdf}), divide by the parameter $\epsilon$, and take the average and average-free parts of the resulting equation to obtain
\begin{subequations}
\begin{align}
\frac{d\tilde{X}_\gamma}{d\theta} &= \langle Q_\gamma (\tilde{X}_\delta + \epsilon Y_\delta, \phi) \rangle \,,
\label{eq2:aveq}\\
\frac{\partial Y_\gamma}{\partial \phi} &= {\cal AF} \left (Q_\gamma (\tilde{X}_\delta + \epsilon Y_\delta, \phi) \right )  - \epsilon \frac{\partial Y_\gamma}{\partial \tilde{X}_\delta} \frac{d\tilde{X}_\delta}{d\theta} \,.
\label{eq2:avfreeeq}
\end{align}
\label{eq2:maineq}
\end{subequations}
These equations can then be iterated in a straightforward way.  At zeroth order, Eq.\ (\ref{eq2:aveq}) yields $d\tilde{X}_\gamma/d\theta = \langle Q^0_\gamma  \rangle$ where $Q^0_\gamma \equiv Q_\gamma (\tilde{X}_\delta, \phi)$, which is the conventional result whereby one averages the perturbation holding the orbit elements fixed.  We define the expansion $Y_\gamma  \equiv Y^{(0)}_\gamma + \epsilon Y^{(1)}_\gamma + \epsilon^2 Y^{(2)}_\gamma + \dots$.  We then integrate Eq.\ (\ref{eq2:avfreeeq}) holding $\theta$ fixed to obtain $Y^0_\gamma$.   The iteration continues until one obtains all contributions to  $d\tilde{X}_\gamma/d\theta$ compatible with the order in $\epsilon$ to which $Q_\gamma$ is known.   The final solution including periodic terms is given by Eq.\ (\ref{eq2:ansatz}), with the secular evolution of the $\tilde{X}_\gamma$ given by solutions of Eqs.\ (\ref{eq2:aveq}).  From these solutions one can reconstruct the instantaneous orbit using Eqs.\ (\ref{eq2:keplerorbit}).  

Although we are interested in the effects of radiation reaction, we must include the conservative terms in the equations of motion.  This is because, for example, a conservative 1PN contribution to $Y$ substituted back into the 2.5PN contribution to $Q$ will lead to a 3.5PN term, as will a 2.5PN contribution to $Y$ substituted back into the 1PN contribution to $Q$.  These must be added to the straight 3.5PN contribution to $Q$. When working to higher orders in the expansion parameter, these so-called ``cross-term'' effects cannot be neglected (see  \cite{2004PhRvD..69j4021M,2017PhRvD..95f4003W} for examples in post-Newtonian theory, in Newtonian triple system dynamics \cite{2016MNRAS.458.3060L,2021PhRvD.103f3003W}, and in triple system dynamics with PN corrections \cite{2014PhRvD..89d4043W,2020PhRvD.102f4033L}).  

\subsection{Results}
\label{sec:results}

We carry out this procedure on the conservative (to 3PN) plus radiation-reaction (to 4.5PN) terms in the equations of motion  (\ref{eq:EOMGen}) to obtain secular evolution equations for $\tilde{p},\ \tilde{\alpha},$ and $\tilde{\beta}$ in terms of the orbital phase $\theta$. The evolution equations resulting from the tail term are derived separately  in Appendix \ref{app:tails}.
Using Eq.\ (\ref{eq:alphabeta}) we convert from $d\tilde{\alpha}/d\theta$ and $d\tilde{\beta}/d\theta$ to $d\tilde{e}/d\theta$ and $d\tilde{\omega}/d\theta$.  We then rescale $p$ by $Gm/c^2$ by defining
\begin{equation}
x \equiv \frac{c^2 p}{Gm} \,,
\label{eq:rescale}
\end{equation}
and obtain the set of equations
\begin{widetext}
\begin{subequations}
\begin{align}
	\frac{d{e}}{d\theta}=
&-\frac{1}{15} \eta e x^{-5/2}  \left(304+121e^2\right)
\nonumber \\
&
+\frac{1}{30} \eta e  x^{-7/2}\left[\frac{1}{28}\left(144392-34768e^2-2251e^4\right)
            +\left(1272-1829e^2-538e^4 \right)\eta \right]
\nonumber \\
&
-\frac{1}{34560} \eta \pi e x^{-4} \left(4538880+6876288 e^2+581208 e^4+623 e^6 \right) 
 \nonumber\\
&
- \frac{1}{120}\eta e x^{-9/2} \biggl [\frac{1}{252}\left(43837360+4258932e^2-1211290e^4+77535e^6\right)\nonumber\\
			&  +\frac{1}{14}\left(1239608-3232202e^2+898433e^4+13130e^6\right)\eta 			 -\left(9216+24353e^2+45704e^4+4304e^6\right)\eta^2
					\biggr ] \,,
\\
\frac{d{x}}{d\theta}=
&-\frac{8}{5} \eta  x^{-3/2}\left(8+7e^2\right)
\nonumber\\
&
+ \frac{1}{15} \eta x^{-5/2}\left [ \frac{1}{14}\left(22072-6064e^2-1483e^4\right) 
          +4\left(36-127e^2-79e^4\right) \eta \right ] 
\nonumber\\
&
-\frac{1}{360} \eta \pi x^{-3} \left(18432+55872e^2+7056e^4-49e^6\right)
\nonumber \\	
&
-\frac{1}{15} \eta x^{-7/2}
\biggl [ \frac{1}{756}\left(8272600+777972e^2-947991e^4-4743e^6\right)
\nonumber\\
&
 +\frac{1}{84}\left(232328-1581612e^2+598485e^4+6300e^6\right) \eta
	-\left(384+ 1025e^2 + 5276e^4+632e^6\right)\eta^2 \biggr ] \,.
\end{align}
\label{eq:dexdtheta}
\end{subequations}	
\end{widetext}
The terms at odd half powers of $1/x$ are the 2.5, 3.5 and 4.5PN contributions respectively (including the relevant cross-term contributions), while the terms at order $x^{-4}$ and $x^{-3}$ are the 4PN tail contributions to $de/d\theta$ and $dx/d\theta$.  Note that the long timescale phase $\theta$ is time divided by the orbital period.  The conservative terms in the equations of motion leave $x$ and $e$ unchanged, reflecting the conservation of energy and angular momentum in the absence of gravitational radiation.   The conservative terms and the 4PN tail term do induce an advance of the pericenter $\omega$ (the 2.5, 3.5 and 4.5PN terms do not), but this has no bearing on the evolution of $x$ and $e$, so we will not consider the pericenter advance further.

\section{Numerical evolutions and the final eccentricity}
\label{sec:numerical}

We now apply the analytic results obtained in the previous section to the long-term evolution of highly eccentric orbits.   Note that the rescaling of $p$ eliminates the total mass of the binary from the equations, so the evolution of $x$ and $e$ depends only on their initial values $x_i$ and $e_i$ and on $\eta$.    The only place where the mass $m$ of the binary will enter the problem is in the conversion from orbital phase to time via the orbital period, and in determining the endpoint of the integrations.  

\begin{figure}[t]
\begin{center}

\includegraphics[width=3.in]{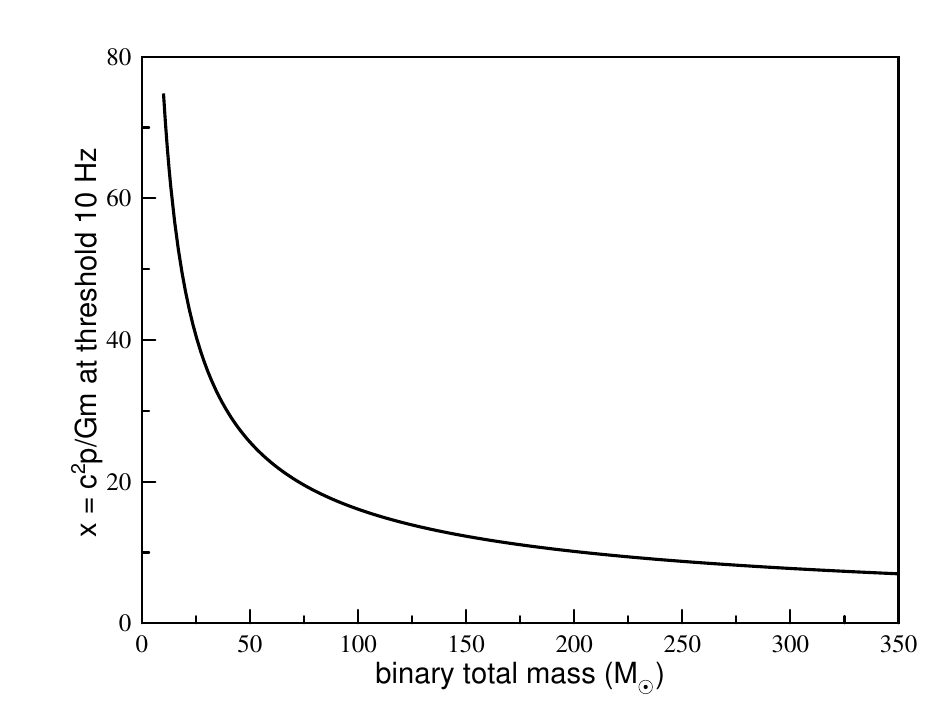}

\caption{\label{fig:pfinvsm} Dependence of $x_f$ on total mass $m$ at the detection threshold of $10$ Hz. }
\end{center}
\end{figure}

Since we are interested in the residual eccentricity in binary inspirals detected via gravitational radiation, we will set that endpoint $x_f$ to be where the gravitational wave frequency corresponds to the frequency where the signal enters the sensitive band of the detector in question.   In this paper we will focus on ground-based detectors, although the results can easily be applied to inspiralling binaries in the LISA band.
For a gravitational wave signal of frequency $f$, and for an orbit of small eccentricity (which is what we expect for the late stage of inspiral), $x_f$ can be approximated by
\begin{align}
x_f &= \left( \frac{c^3}{\pi Gm f} \right)^{2/3} 
=
    47.12  \left ( \frac{20 M_\odot}{m} \frac{10 {\rm Hz}}{f} \right )^{2/3}   \,.
    \label{eq:xf-vs-mass}
\end{align}
Figure \ref{fig:pfinvsm} shows $x_f$ as a function of total mass $m$ for a threshold detection frequency of $10$ Hz.
Note that the values of $x_f $ are sufficiently large that our use of the Newtonian formula for the orbital period $P = 2\pi [p/(1-e^2)]^{2/3}/(Gm)^{1/3}$ is justified, as is the use of the post-Newtonian approximation to analyze the evolutions to this end point.     And as next generation detectors, such as Einstein Telescope and Cosmic Explorer, begin detecting even lower frequency signals (larger $x_f$) the PN approximation for estimating residual eccentricities for a given binary system mass will be even better.  

\begin{figure}[t]
\begin{center}

\includegraphics[width=3.7in]{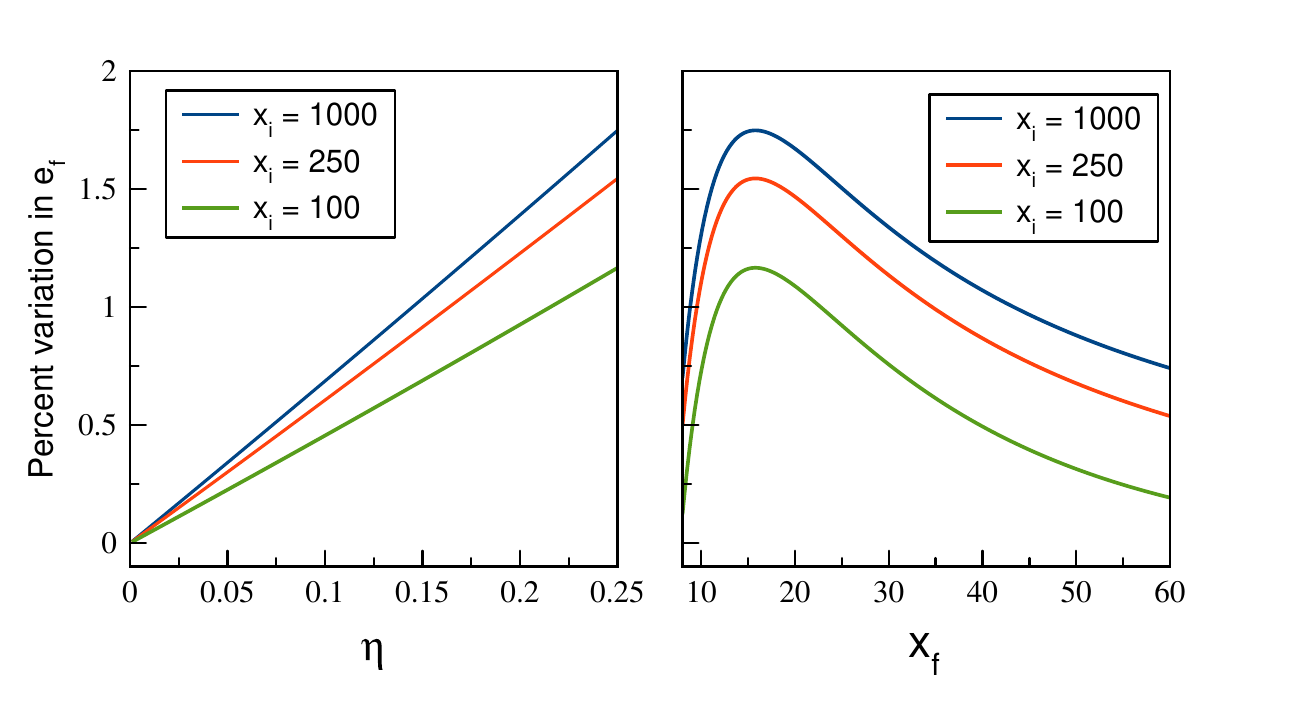}

\caption{\label{fig:etadependence} Dependence of $e_{f}$ on $\eta$.  Left: Plotted is $e_{f}$ normalized by its value for $\eta = 5 \times 10^{-5}$ (effectively zero), vs. $\eta$ for various initial values of $x$ and for $x_{f} = 16$ corresponding to a $100 \, M_\odot$  binary entering the LIGO/Virgo band at $10$ Hz.  Right: Percent difference between $e_f$ for equal-mass and extreme mass-ratio systems, as a function of $x_f$.}
\end{center}
\end{figure}

\subsection{Dependence on $\eta$}
\label{sec:eta}

We first explore the dependence of the final eccentricity on the symmetric mass ratio $\eta$.  Although $dx/d\theta$ and $de/d\theta$ are proportional to $\eta$, that factor cancels in the ratio $dx/de$, so the only dependence on $\eta$ arises through the  various PN correction terms.  Not surprisingly, these effects are small, as can be seen in the left panel of Fig.\ \ref{fig:etadependence}, where we plot the percent variation in  $e_f (\eta)$ relative to $e_f(0)$ as a function of $\eta$, for a selection of $x_i$ and for $x_f$ corresponding to a $100 \, M_\odot$  source entering the LIGO/Virgo band at $10$ Hz.   In the numerical integrations we choose $\eta = 5 \times 10^{-5}$ in lieu of zero.   In the right panel we plot the percent difference between $e_f$ for $\eta =1/4$ and $e_f$ for $\eta = 0$ as a function of $x_f$.   The variations are less than two percent over the ranges of the parameters. 

\begin{figure}[t]
\begin{center}

\includegraphics[width=3.4in]{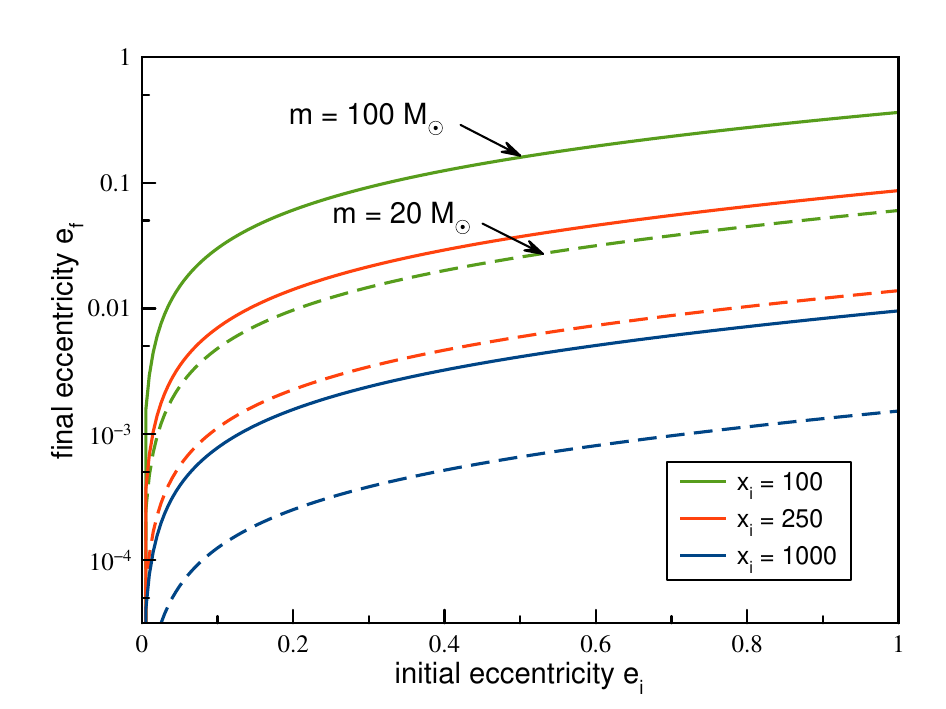}

\caption{\label{fig:efinein} Final eccentricity vs. initial eccentricity.  Solid and dashed lines correspond to total masses of $m$ of $100 M_\odot$ and $20 M_\odot$, respectively.  Green, red and blue curves correspond to $x_i = 100$, $250$ and $1000$, respectively.   }
\end{center}
\end{figure}

Figure \ref{fig:efinmass} shows the
 eccentricities reached at the LIGO/Virgo detection threshold as a function of the total mass of the binary system, for various initial values of $x$ and $e$.   Fig.\ \ref{fig:efinein} shows the final eccentricities as a function of initial eccentricities for various initial values of $x$ and of the total mass. 

\subsection{Accuracy of the PN approximation}
\label{sec:PNconvergence}

We next investigate the accuracy of the PN approximation.  There is no formal way to do this because the PN sequence is not known to be a convergent series (at best, it might be an asymptotic sequence), but one way to estimate the accuracy is to add terms at the next PN order, and to study their effects.  We use the equations of evolution for test-body orbits taken from black hole perturbation theory obtained by Sago and Fujita (SF) \cite{2015PTEP.2015g3E03S}.  After showing that, following a suitable transformation between their definitions of orbit elements and ours (see the discussion in Appendix \ref{app:55PN}), the equations are in agreement through 4.5PN order, we add the 5.5PN terms.  These have the form 
\begin{align}
\frac{de}{d\theta} \biggl |_{\rm 5.5} & = 
\frac{\eta e x^{-11/2}}{349272000}  \bigl(1790315545528
\nonumber \\
& \qquad -6186148025656e^2-4964186588931e^4\bigr) \,,
\nonumber \\
\frac{dx}{d\theta} \biggl |_{\rm 5.5} & = 
\frac{\eta x^{-9/2} }{87318000}  \bigl ( 294262221896-621776393808e^2
\nonumber \\
& \qquad -658790352267e^4
-277665065676e^6\bigr) \,.
\label{eq:55}
\end{align}
 
\begin{figure}[t]
\begin{center}

\includegraphics[width=3.4in]{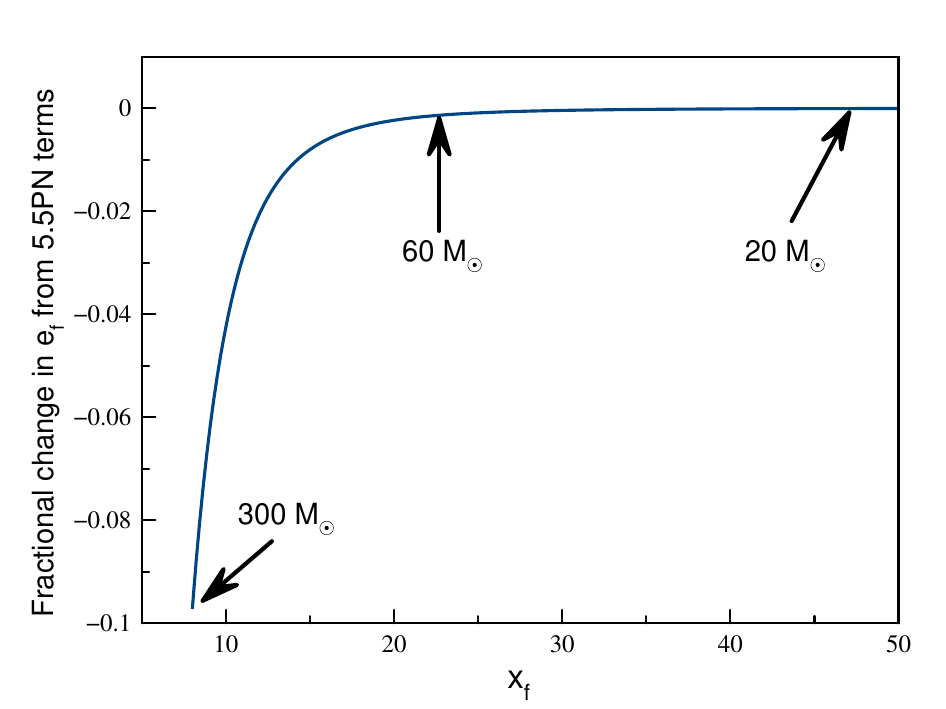}

\caption{\label{fig:55PN} Effect of 5.5PN terms on $e_{f}$ at the LIGO/Virgo threshold, for $e_i = 0.999$ and $x_i = 1000$.  The effect is very small, except for the most massive systems, which are very relativistic when they cross the threshold.   }
\end{center}
\end{figure}

We add these terms directly to Eqs.\ (\ref{eq:dexdtheta}), without worrying about ``PN cross term" contributions at this order, and carry out various numerical integrations.  A typical result is plotted in Fig.\ \ref{fig:55PN}, showing the fractional difference $e_f(5.5PN)/e_f(4.5PN) -1$ for $e_i = 0.999$ and $x_i = 1000$ as a function of $x_f$.  The effect is less than one percent over a wide range of $x_f$, reaching 10 percent only for very massive sources entering the LIGO/Virgo band in a highly relativistic regime, $x_f \sim 8$ ($m \sim 300 M_\odot$).      

\begin{figure}[t]
\begin{center}

\includegraphics[width=3.4in]{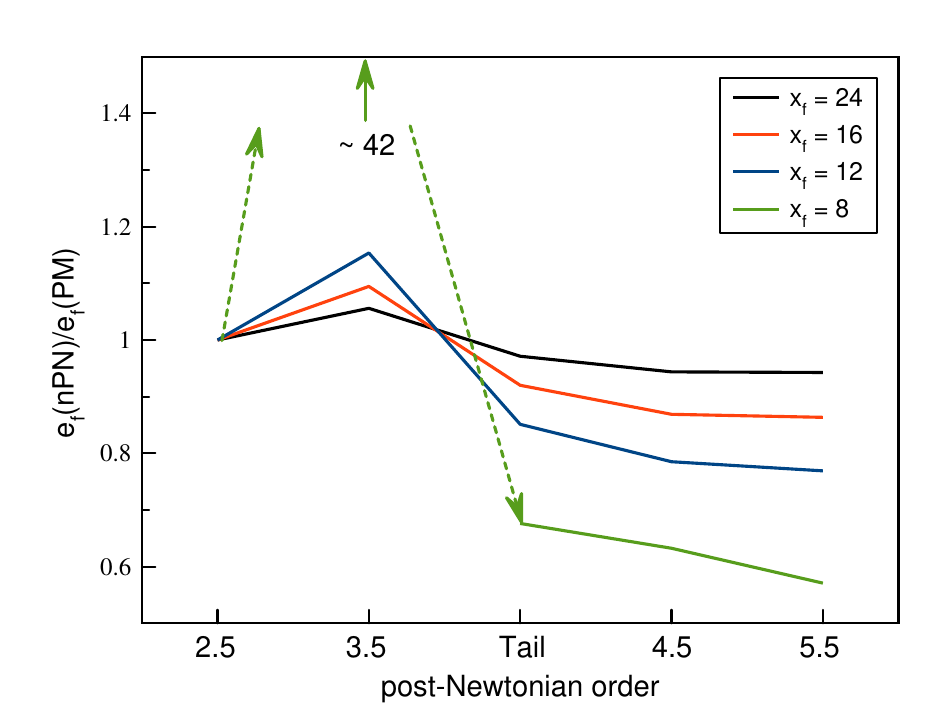}

\caption{\label{fig:PNconvergence} Effect of PN corrections on values of $e_f$ as compared to the Peters-Mathews map. At 2.5PN order the numerical results agree with the PM values.  At 3.5PN order, the sign difference between the 3.5PN and 2.5PN terms causes $e$ to begin to grow when $x \approx 8.5$, leading to $e_f$ at $x_f = 8$ around 42 times the PM value.  Adding additional PN terms mitigates this singular behavior, leading to values that behave ``nicely'' as PN orders are added.   }
\end{center}
\end{figure}

Another way to explore the behavior of the PN approximation is to examine the values of $e_f$ as a function of the post-Newtonian order of the equations of evolution (\ref{eq:dexdtheta}).   At the lowest 2.5PN order, our numerical results for $e_f$ as a function of $x_f$ are in close agreement with those obtained from the  Peters-Mathews map, Eq. (\ref{eq:PM}). 
In Fig.\ \ref{fig:PNconvergence} we plot our numerical results for $e_f$ normalized by the values from the PM formula, when the various PN corrections are added one by one to the system of equations.  The initial conditions in all the examples are $e_i = 0.999$ and $x_i = 250$; the final values of $x$ range from $8$ to $24$.  The first conclusion is that in all cases, the results seem to ``converge", in the sense of behaving ``nicely'' as PN orders are added.  Needless to say, demonstrating actual convergence is essentially impossible.  The other notable feature is that the first PN corrections to the leading terms, i.e.\ the 3.5PN corrections, have the effect of increasing the value of $e_f$ relative to the PM value.  This is because of the relative negative sign between the 2.5PN and the 3.5PN terms in Eqs.\ (\ref{eq:dexdtheta}).   In fact, for the most relativistic case of $x_f = 8$ the 3.5PN terms actually {\em dominate} the 2.5PN terms when $x$ reaches about $8.5$, leading to a {\rm growing} eccentricity, eventually producing a final eccentricity  42 times larger than the PM value!   Adding the 4PN tail term, the 4.5PN term and the 5.5PN term mitigates this behavior, leading to a well-behaved sequence of values for $e_f$, even in the very relativistic regime.  In all cases the PN-corrected values for $e_f$ are {\em smaller} than the Peters-Mathews values.  In the next subsection we search for an analytic extension of the PM formula that accurately encapsulates these post-Newtonian corrections.

\subsection{A PN-corrected map for $e$}

Our goal is to obtain an analytic map for the eccentricity of a generic binary inspiral at a given value of $x$ that extends the Peters-Mathews map (\ref{eq:PM}) into the relativistic regime.   Our criteria are that the map be trivially satisfied when $e = e_i$ and $x = x_i$, that in the limit of large $x_i$ and $x$ it tend to the PM map, and finally that it differ from the numerical values by only a few percent, down to $x_f \sim 8$.  After some experimentation, we arrived at Eq.\ (\ref{eq:PNcorrected}).  Formally, $e$ will then be given by
\begin{equation}
e = g^{-1} \left [ \frac{x}{x_i} \left ( \frac{1+2/x}{1+2/x_i} \right ) \left 
( \frac{1-4/x}{1-4/x_i} \right )^{12/19} g(e_i) \right ] \,.
\label{eq:PNmap}
\end{equation}
In practice, of course, Eq.\ (\ref{eq:PNmap}) must be solved numerically for $e$.   If one prefers to work in terms of semimajor axis rather than semilatus rectum, the conversion is $x = a(c^2/Gm)(1-e^2)$.

\subsection{Evolution time}

In order to determine the probability of a given final eccentricity arising from astrophysically meaningful initial conditions, it is important also to know the time for the system to evolve from the initial state to the final state.  That time can be found by numerically integrating the system of equations $dx/dt = (2\pi/P) dx/d\theta $ and $de/dt = (2\pi/P)  de/d\theta $, where $P$ is the orbital period.  Because the total time is strongly dominated by the non-relativistic part of the orbit, it suffices to use the Keplerian orbital period, $P = 2\pi (Gm/c^3) x^{3/2} (1-e^2)^{-3/2}$.  At the lowest, 2.5PN order of approximation, Eqs.\ (\ref{eq:dexdtheta}) yield
\begin{equation}
\frac{dx}{dt} = - \frac{64}{5} \eta \frac{c^3}{Gm} x^{-3} (1-e^2)^{3/2} \left ( 1 + \frac{7}{8} e^2 \right )\,.
\label{eq:dxdt}
\end{equation}
In the circular limit ($e=0$), integrating this equation directly yields
\begin{equation}
T = \frac{5}{256 \eta} \left ( \frac{Gm}{c^3} \right ) x_i^4 \left ( 1- \frac{x_f^4}{x_i^4} \right ) \,.
\label{eq:Tcirc}
\end{equation}
If the initial eccentricity is large, most of the time will be spent in  the large $e$ regime.  From Eqs.\ (\ref{eq:dexdtheta}), again at 2.5PN order, we can thus approximate $dx/de \approx (72/85) x/e$, and thus $e \approx e_i (x/x_i)^{85/72}$.  We then obtain
\begin{equation}
\frac{dz}{dt} \approx -24 \eta  \frac{c^3}{Gm} x_i^{-4} \frac{(1-e_i^2 z^{85/36} )^{3/2} }{z^3} \,,
\end{equation}
 where $z \equiv x/x_i$.  Because of the very weak dependence of the time on $x_f$ when $x_f \ll x_i$ (see Eq.\ (\ref{eq:Tcirc})), we can integrate this equation from $z=0$ to $z = 1$, with the result
\begin{equation}
T = \frac{1}{96 \eta} \left ( \frac{Gm}{c^3} \right ) x_i^4 G(e_i) \,,
\label{eq:Tecc}
\end{equation}
where
\begin{align}
G(e_i) &= 4 \int_0^1 \frac{z^3 dz}{(1-e_i^2 z^{85/36} )^{3/2}}
\nonumber \\
& = \, _2F_1 \left (\frac{3}{2},\frac{144}{85};\frac{229}{85};e_i^2 \right )
\nonumber \\
& \approx \frac{ 288/85}{\sqrt{1-e_i^2}} - 5.9257 + 8.0919 \sqrt{1-e_i^2} + \dots \,,
\end{align} 
where $ _2F_1 (a,b;c;z)$ is the hypergeometric function, and the final expression is the expansion for $e_i \approx 1$ \cite{1964PhRv..136.1224P,2009MNRAS.395.2127O,2017PhRvD..95f4003W}.   The singular behavior of $G(e_i)$ merely reflects the enormous amount of time bodies in eccentric orbits spend going to apocenter and back, with no relativistic consequences to speak of.

\begin{figure}[t]
\begin{center}

\includegraphics[width=3.7in]{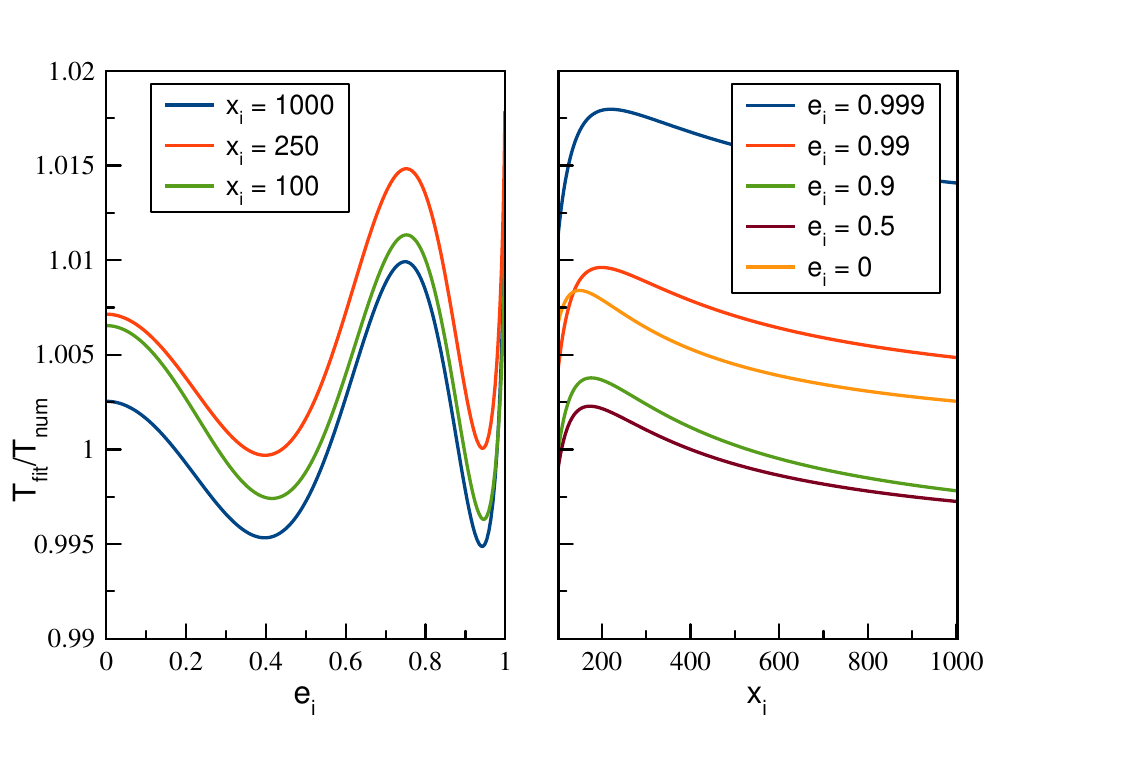}

\caption{\label{fig:Tfits} Comparison between the analytic fit for $T$ and the numerical values, plotted vs. $e_i$ and $x_i$.  }
\end{center}
\end{figure}

Our goal then is to combine Eq.\ (\ref{eq:Tcirc}) with Eq.\ (\ref{eq:Tecc}), and to tweak the result to obtain a decent fit to the numerical values for the time $T$. After some experimentation, we obtain the (less than elegant) result
\begin{align}
T_{\rm fit} &=  \frac{5}{256 \eta} \left ( \frac{Gm}{c^3} \right ) \left ( {x_i^4-x_f^4} \right )
\left ( 1+ \frac{8(1-e_i^2)}{x_i} \right ) {\cal G}(e_i) 
\nonumber \\
&=  0.244 \,\left (\frac{1}{4\eta}\right ) \left (\frac{m}{20 M_\odot}\right ) \left ( \frac{x_i}{1000} \right )^4  \, {\rm yr} 
\nonumber \\
& \quad \times  \left ( 1 -\frac{x_f^4}{x_i^4} \right )
\left ( 1+ \frac{8(1-e_i^2)}{x_i} \right ) {\cal G}(e_i) \,,
\end{align}
where
\begin{align}
{\cal G}(e_i) &= 1 + \frac{288}{85} \left ( \frac{1-\frac{19}{40} e_i^4 }{\sqrt{1-e_i^2}}-1 \right ) 
\nonumber \\
& \qquad 
- \frac{5}{8} e_i^2 (1-e_i^2)(2+e_i^2+e_i^4 ) \,.
\end{align}
This formula agrees with the numerically generated values of $T$ to better than two percent over the range $100 \le x_i \le 1000$ and $0 \le e_i \le 0.999$ (see Fig.\ \ref{fig:Tfits}).

\section{Discussion and Conclusions}
\label{sec:conclusions}

We have used post-Newtonian equations of motion containing radiation-reaction terms through 4.5PN order to analyse the late-time eccentricities of inspiraling binary systems of non-spinning compact bodies of arbitrary masses.  We have found that, apart from the overall dependence of the inspiral time $T$ on the symmetric mass ratio $\eta$, the final eccentricities are essentially independent of $\eta$.  We found an analytic map for the final eccentricity in terms of the initial eccentricity and semi-latus rectum that generalizes the Peters-Mathews formula, and that agrees with the numerically generated values to a few percent.   We also find that the Peters-Mathews formula produces consistently larger values, by as much as 60 percent, than those predicted by the full 4.5PN dynamics.  These results may be useful for assessing the levels of orbital eccentricity that must be incorporated into gravitational-waveform templates, and for relating measured late-time eccentricities to the astrophysical origins of compact binary inspirals.

Here we note that, in the limit $e_i \ll 1$, Tanay et al.\ \cite{2016PhRvD..93f4031T} obtained a PN-corrected map by adding post-Newtonian corrections to the {\em inversion} of the PM map (\ref{eq:PM}) expressed as a power series in $e_i$  \cite{2009PhRvD..80h4001Y}.

\acknowledgments

This work was supported in part by the National Science Foundation,
Grants No.\ PHY 16-00188 and PHY 19-09247.   We are grateful for the hospitality of  the Institut d'Astrophysique de Paris where part of this work was carried out.  We also thank Imre Bartos for useful discussions at the early stage of this work.


\appendix


\section{Coefficients in the PN equations of motion}
\label{app:coeffs}

Here we list the coefficients $a^{(N)}_{lmn}$, $b^{(N)}_{lmn}$, $c^{(N)}_{lmn}$, and $d^{(N)}_{lmn}$ that appear in the conservative and radiation-reaction parts of the equations of motion (\ref{eq:EOMGen}).  See \cite{2014LRR....17....2B} for a review of the conservative contributions in Table \ref{tab:conservative}.  The 3.5PN terms in Table \ref{tab:rr}  are taken from \cite{2002PhRvD..65j4008P}.
The 4.5PN terms are taken from \cite{1997PhRvD..55.6030G}; because the terms were derived from an energy and angular momentum balance argument, there are 12 arbitrary gauge-type parameters (Table \ref{tab:gauge}).  Those parameters disappear in the orbit-averaged equations for the orbit elements.

\begin{table*}[ht]
\caption{\label{tab:conservative} Coefficients in conservative terms}
\begin{tabular}{lcc@{\hskip 0.5cm}c}
 \hline
$\{l,m,n\}$	&& $a^{(N)}_{lmn}$ 	&$b^{(N)}_{lmn}$ \\[1ex]
 \hline \hline
  \multicolumn{4}{c}{1 PN Order ($N=1$)} \\
  \hline
$\{1,0,0\}$ 	&& $2(2+\eta)$   		& $0$\\[1ex]
$\{0,1,0\}$		&&  $\frac{3}{2}\eta$ 		& $2(2-\eta)$ \\[1ex]
$\{0,0,1\}$		&& $-(1+3\eta)$ 			& $0$ \\ [1ex]
 \hline
  \multicolumn{4}{c}{2 PN Order ($N=2$)} \\
 \hline
 $\{2,0,0\}$		&& $-\frac{3}{4}(12+29\eta)$			& $0$\\[1ex]
$\{0,2,0\}$		&& $-\frac{15}{8}\eta(1-3\eta)$ 		& $-\frac{3}{2}\eta(3+2\eta)$ \\[1ex]				
$\{0,0,2\}$		&& $-\eta(3-4\eta)$ 					& $0$ \\ [1ex] 	
$\{1,1,0\}$		&& $2+25\eta+2\eta^2$ 				& $-\frac{1}{2} (4+41\eta+8\eta^2)$  \\ [1ex]	
$\{1,0,1\}$		&& $\frac{1}{2}\eta(13-4\eta)$ 		& $0$ \\ [1ex]
$\{0,1,1\}$		&& $\frac{3}{2}\eta(3-4\eta)$ 		& $\frac{1}{2}\eta(15+4\eta)$ \\ [1ex] 	
 \hline 
  \multicolumn{4}{c}{3 PN Order ($N=3$)} \\
  \hline
$\{3,0,0\}$	&&{$16+ \frac{1}{48}\eta(5596-123\pi^2+1704 \eta)$}  	&{$0$}\\[1ex]	
$\{0,3,0\}$	&&{$\frac{35}{16}\eta(1-5\eta+5\eta^2)$ }	& {$\frac{15}{8}\eta(3-8\eta-2\eta^2)$}\\[1ex]	 
$\{0,0,3\}$	&&{$-\frac{1}{4}\eta(11-49\eta+52\eta^2)$} 		&{$0$}\\ [1ex]	
$\{2,1,0\}$	&& {$-1-\left(\frac{22717}{168}+\frac{615}{64}\pi^2\right)\eta-\frac{11}{8}\eta^2+7\eta^3$}& 
{$4+\left(\frac{5849}{840}+\frac{123}{32}\pi ^2 \right )\eta -25\eta^2-8 \eta^3 $}\\ [1ex]
$\{2,0,1\}$	&& {$\eta\left(\frac{20827}{840}+\frac{123}{64}\pi^2-\eta^2\right)$} 	& {$0$}\\ [1ex]
$\{1,2,0\}$	&& {$-\frac{1}{2}\eta(158-69\eta-60\eta^2)$} 												&{$-\frac{1}{6}\eta(329+177\eta+108\eta^2)$}\\ [1ex]
$\{1,1,1\}$	&&{$\eta(121-16\eta-20\eta^2)$ }															&{$\eta(15+27\eta +10\eta^2)$}\\ [1ex]
$\{1,0,2\}$	&& {$-\frac{1}{4}\eta(75+32\eta-40\eta^2)$ }													& {$0$}\\[1ex]
$\{0,2,1\}$	&& {$-\frac{15}{8}\eta(4-18\eta+17\eta^2)$} 												& {$-\frac{3}{4}\eta(16-37\eta-16\eta^2)$}\\ [1ex]
$\{0,1,2\}$	&& {$\frac{3}{8}\eta(20-79\eta+60\eta^2)$} 													& {$\frac{1}{8}\eta(65-152\eta-48\eta^2)$}\\ [1ex]
\hline
 \end{tabular}
 \end{table*}
%
 %
 \begin{table*}[ht]
\caption{\label{tab:rr} Coefficients in radiation reaction terms. (In \cite{2017PhRvD..95f4003W}, the first term in the parentheses in $c^{(3)}_{0,1,2}$ was incorrectly given as $295$.) }
\begin{tabular}{lcc@{\hskip 0.5cm}c}
 \hline
$\{l,m,n\}$	&& $c^{(N)}_{lmn}$ 	&$d^{(N)}_{lmn}$ \\[1ex]
 \hline\hline
  \multicolumn{4}{c}{2.5 PN Order ($N=1$)} \\
  \hline
 $\{1,0,0\}$	&& $\frac{17}{3}$ & $-3$\\[1ex]
  $\{0,1,0\}$	&& $0$ 				& $0$ \\[1ex]
$\{0,0,1\}$		&&  $3$ 	& $-1$\\ [1ex]
   \hline
  \multicolumn{4}{c}{3.5 PN Order ($N=2$)} \\
  \hline
 $\{2,0,0\}$		&&  $-\frac{23}{14}(43+14\eta)$		& $\frac{1}{42}(1325+546\eta)$\\[1ex]
$\{0,2,0\}$		& &  $-70$ 							& $75$ \\[1ex]					
$\{0,0,2\}$		& &  $-\frac{3}{28}(61+70\eta)$ 		& $\frac{1}{28}(313+42\eta)$\\ [1ex] 	
$\{1,1,0\}$		& &  $-\frac{1}{4}(147+188\eta)$ 		& $\frac{1}{12}(205+424\eta)$\\ [1ex]
$\{1,0,1\}$		& &  $-\frac{1}{42}(519-1267\eta)$ 	& $-\frac{1}{42}(205+777\eta)$\\ [1ex]
$\{0,1,1\}$		& &  $\frac{15}{4}(19+2\eta)$ 					& $-\frac{3}{4}(113+2\eta)$\\ [1ex] 	
   \hline 
  \multicolumn{4}{c}{4.5 PN Order ($N=3$)} \\
  \hline
$\{3,0,0\}$	&&{$\frac{1}{756}(289079+284127 \eta+22632\eta^2)+\mathcal{C}_{300} $}	& {$-\frac{1}{2268} (395929+398700 \eta+ 87048 \eta^2)+\mathcal{D}_{300}$}\\[1ex]	
$\{0,3,0\}$	&& {$\mathcal{C}_{030}$}																		& {$\frac{5}{18}(291-919 \eta +97\eta^2)+\mathcal{D}_{030}$}\\[1ex]			
$\{0,0,3\}$	&& {$\frac{1}{168}(779+604\eta-7090\eta^2)+\mathcal{C}_{003}$ }									& {$-\frac{1}{56}(834-1956 \eta- 1743\eta^2)+\mathcal{D}_{003}$}\\ [1ex]	
$\{2,1,0\}$	&& {$\frac{1}{756}(250221-6032\eta+74134\eta^2)+\mathcal{C}_{210}$}		& {$-\frac{1}{252} (37992+62832\eta+9649\eta^2) +\mathcal{D}_{210}$}\\ [1ex]
$\{2,0,1\}$	&& {$-\frac{1}{252}\left(20916-24324 \eta+23483\eta^2 \right )+\mathcal{C}_{201}$} 		& {$\frac{1}{252}(26703+21304\eta+28486\eta^2)+\mathcal{D}_{201}$}\\ [1ex]
$\{1,2,0\}$	&& {$\frac{1}{252}(108322-43996\eta+12839\eta^2)+\mathcal{C}_{120}$} 	& {$-\frac{1}{252}(99499+24002\eta +33443\eta^2)+\mathcal{D}_{120}$}\\ [1ex]
$\{1,1,1\}$	&& {$-\frac{1}{504}(218401-160227\eta +95987\eta^2)+\mathcal{C}_{111}$} 	& {$\frac{1}{504}(200244+65460 \eta + 83501\eta^2)+\mathcal{D}_{111}$}\\ [1ex]
$\{1,0,2\}$	&& {$\frac{1}{504}(40758-88311\eta+43474\eta^2)+\mathcal{C}_{102}$} 	& {$-\frac{1}{504}(16731+24785 \eta+41471\eta^2)+\mathcal{D}_{102}$}\\[1ex]
$\{0,2,1\}$	&& {$\frac{5}{18}(87-215\eta - 97\eta^2)+\mathcal{C}_{021}$} 				
& {$-\frac{5}{168}(6889-21631\eta+2380\eta^2)+\mathcal{D}_{021}$}\\ [1ex]
$\{0,1,2\}$	&& {$-\frac{1}{84}(1205-260 \eta -8785\eta^2)+\mathcal{C}_{012}$ }						& {$\frac{1}{168} (21280-60733\eta-11999\eta^2)+\mathcal{D}_{012}$}\\ [1ex]
\hline
 \end{tabular}
 \end{table*}

\begin{table}[ht]
\caption{\label{tab:gauge} 4.5PN gauge coefficients}

\begin{tabular}{lll@{\hskip 0.5cm}l}
 \hline
 $\{l,m,n\}$&& $\mathcal{C}_{lmn}$ 			&$\mathcal{D}_{lmn}$\\[1ex]
 \hline
$\{3,0,0\}$	&& $-2\psi_5-3\psi_9$   			& $-\psi_5$\\[1ex]	
$\{0,3,0\}$	&& $-9 \psi_7$ 		& $-7\chi_8$\\[1ex]			
$\{0,0,3\}$	&& $3(\psi_2 - \chi_6)$ 			& $\psi_1$\\ [1ex]		
$\{2,1,0\}$	&& $-2\psi_6-5\psi_8-7\psi_9$ 		& $-3\chi_9-2\psi_3-5\psi_5$ \\ [1ex]
$\{2,0,1\}$	&& $3\chi_9-2\psi_3-3\psi_6+3\psi_9$ 		& $\psi_5-\psi_3$\\ [1ex]
$\{1,2,0\}$	&& $-2\psi_4-7\psi_7-8\psi_8$ 		& $-2\chi_6-5\chi_8-6\chi_9$\\ [1ex]
$\{1,1,1\}$	&& $2\chi_6+5\chi_8+6\chi_9-4\psi_2$ 	
& $3\chi_9-3\chi_6-4\psi_1$\\ [1ex]
&&$\quad -5\psi_4-6\psi_6+5\psi_8$&$\quad-4\psi_3$\\[1ex]
$\{1,0,2\}$	&& $3\chi_6-3\chi_9-2\psi_1$ 		& $-\psi_1+\psi_3$\\[1ex]
&&$\quad -3\psi_2+3\psi_6$&\\[1ex]
$\{0,2,1\}$	&& $7(\chi_8-\psi_4+\psi_7)$ 					& $5(\chi_8-\chi_6)$\\ [1ex]
$\{0,1,2\}$	&& $5(\chi_6-\chi_8-\psi_2+\psi_4)$ 				& $3(\chi_6-\psi_1)$\\ [1ex]
\hline
\end{tabular}
\end{table}
 

 \section{4PN tail terms}
 \label{app:tails}
 
In this Appendix, we derive the effects of the leading 4PN gravitational-wave ``tail'' term on the evolution of $e$ and $x$.  Our starting point is Eq.\ (\ref{eq:tailterm}) with the trace-free quadrupole moment given by Eq.\ (\ref{eq:quadmom}).

The seven time derivatives of $\mathcal I ^{\left<jk\right>}$ in terms of $v,\ m/r$ and $\dot{r}$ can be easily calculated by iteratively applying the Newtonian equations of motion
\begin{align}
	\frac{dr_i}{dt}&=v_i \,,
	\nonumber \\
	\frac{dv_i}{dt}&= -\frac{Gm}{r^3}r_i \,,
	\nonumber \\
	\frac{d\dot{r}}{dt} & = \frac{1}{r}\left(v^2 -\dot{r}^2 - \frac{Gm}{r}\right) \,,
	\nonumber \\
	\frac{d v^2}{dt}&=-2\frac{Gm}{r^2}\dot{r} \,.
\end{align}  

 To find the secular evolution of the orbital parameters, we will need to integrate over the past history of the binary system, and the structure of that integral dictates that we work in terms of time. Instead of using the true anomaly $f$, we use  the eccentric anomaly $u$, which has a simple, if transcendental relation to time via Kepler's equation
\begin{align}
	t & = \sqrt{\frac{a^3}{G m}}(u-e \sin u) \,,
	\label{eq:Kep}
\end{align}
where $a$ is the semi-major axis, together with the relations
\begin{equation}
	\cos f = \frac{\cos u-e}{1-e \cos u} \,,
	\quad \sin f = \frac{\sqrt{1-e^2}\sin u}{1-e\cos u} \,.
	\label{eq:anomalyconv}
\end{equation} 

 A common way to invert Eq.\ (\ref{eq:Kep}) is to expand in powers of $e < 1$; in order to be compatible with the orders of our non-tail expressions, we will expand to order $e^6$ :
\begin{align}
	u = \tilde{t} + \sum_{i=1}^6{l_i e^i} \,,
	\label{eq:aexp}
\end{align}
where $\tilde{t}=t\sqrt{Gm/a^3}$.
Substituting Eq.\ (\ref{eq:aexp}) into Eq.\ (\ref{eq:Kep}), expanding to ${O}(e)^6$ and demanding that the equality hold order by order, allows us to solve for the coefficients $l_i$. Substituting those expressions into Eq.\ (\ref{eq:aexp}) gives the relation
\begin{align}
u(t) =& \tilde{t}+e \sin \tilde{t}+\frac{1}{2} e^2 \sin 2 \tilde{t} -\frac{1}{8} e^3 (\sin \tilde{t}-3 \sin 3 \tilde{t})
\nonumber \\
&
 -\frac{1}{6} e^4 (\sin 2 \tilde{t} - 2 \sin 4 \tilde{t})\nonumber \\
	&
 +\frac{1}{384} e^5 (2 \sin \tilde{t}-81 \sin 3 \tilde{t}+125 \sin 5 \tilde{t})
\nonumber \\
& +\frac{1}{240} e^6 (5 \sin 2 \tilde{t}-64 \sin 4 \tilde{t}+81 \sin 6 \tilde{t}) \,,
	\label{eq:ut}
\end{align}

We now follow a similar procedure to that detailed in Sec.\ \ref{sec:evolution} B, but using the eccentric anomaly instead of the true anomaly.
We place the orbit on the $X-Y$ plane and use the orbital equations
\begin{align}
	r  &\equiv  a(1-e\cos u)\,,
	\nonumber \\
	\bm{n} & \equiv\frac{\cos u - e}{1 - e \cos u} \bm{e}_X
	+\frac{\sqrt{1-e^2}\sin u}{1-e\cos u}\bm{e}_Y \,,
	\nonumber \\
	\dot{r} & \equiv  \sqrt{\frac{Gm}{a}}\frac{e \sin u}{1-e\cos u} \,, 
	\nonumber \\
	v^2 & \equiv \frac{Gm}{a} \,\frac{1+e\cos u}{1-e \cos u} \,.
\end{align}
We then compute the radial $\mathcal{R}\equiv \bm{n} \cdot \bm{a}_{\rm Tail}$, and cross-track $\mathcal{S}\equiv \bm{\lambda} \cdot \bm{a}_{\rm Tail}$ components of the perturbing acceleration ($\mathcal{W}=0$) and insert these into the Lagrange planetary equations expressed in terms of $t$ and $u$:
\begin{align}
	\frac{dp}{dt}&=2\sqrt{\frac{p^3}{Gm}} \frac{(1-e \cos u)}{1-e^2} \mathcal{S} \,,
	\nonumber \\
	\frac{de}{dt}& = \sqrt{\frac{p}{Gm}}\frac{1}{1-e\cos u}\bigl [ (1-e^2)^{1/2}\sin u \, \mathcal{R}
	\nonumber \\
	& \quad +\left(2\cos u-e(1+\cos^2 u) \right)\mathcal{S} \bigr ] \,.
	\label{eq:tailLPE}
\end{align}

Substituting Eq.\ (\ref{eq:ut}) into the the full expressions for Eqs.\ (\ref{eq:tailLPE}) and being careful to expand to the appropriate order of $e$, we make the transformation $\tilde{t}\rightarrow \tilde{t}-\tilde{s}$ in the evaluated seventh time derivative of ${{\cal I}^{\langle jk \rangle}}$. We average over one orbit with $\tilde{t}$ as the variable of integration from $0$ to $2\pi$. This gives us the secular evolution equations for $dp/dt$ and $de/dt$. To carry out the remaining integral over $\tilde{s}$ into the infinite past we make use of the well known results:
\begin{align}
	\int_0^\infty \cos (py) \ln(y)dy &= -\frac{\ln(p)+\gamma}{p}, \nonumber \\
	\int_0^\infty \sin(py) \ln(y) dy &= -\frac{\pi}{2p} \,,
\end{align}
where $\gamma$ is the Euler number
 (see Eqs.\ 4.441 in  \cite{GradshteynRyzhik}). We then rescale $p$ using Eq.\ (\ref{eq:rescale}) and convert from $a$ to $x$ with $a=(Gm/c^2)x/(1-e^2)$. We convert from $d/dt$ to $d/d\theta$ with $dF/dt = (2\pi/P) dF/d\theta$ where we only need to use the Keplerian expression for the orbital period $P=2\pi(Gm/c^3)x^{3/2}(1-e^2)^{-3/2}$.  We arrive finally at the terms in Eqs.\ (\ref{eq:dexdtheta}) that are of the order $x^{-4}$ in $de/d\theta$ and $x^{-3}$ in $dx/d\theta$.  Because the tail effects are already of 4PN order, we do not have to worry about ``cross terms'' in the two-timescale analysis.
 
  \section{5.5PN terms}
 \label{app:55PN}
 
Here we derive the 5.5PN contributions to $dx/dt$ and $de/dt$ shown in Eqs.\ (\ref{eq:55}).   We combine expressions for $E$ and $L$ to 3PN order in harmonic coordinates explicitly given by Will and Maitra (WM) in Eqs.\ (3.24) in \cite{2017PhRvD..95f4003W} with expressions for the energy and angular momentum flux to the same 3PN order, calculated to $O(e^6)$ by Sago and Fujita \cite{2015PTEP.2015g3E03S} for a point particle orbiting a Kerr black hole using Boyer-Lindquist (BL) coordinates. For our purposes, we take the spin to be zero. The Sago-Fujita (SF) results are given in their Eqs.\ (32) and (33).

First we  must translate from SF's choice of Boyer-Lindquist orbital variables $(e_{\mathrm{BL}},\ p_{\mathrm{BL}})$ to osculating orbit elements in harmonic coordinates $(e,\ p)$, consistent with our work. We do this by employing the method detailed in \cite{2019CQGra..36k5001T}, in which one defines an ``invariant'' dimensionless semilatus rectum $x_0$ and eccentricity $e_0$ via the gauge invariant energy and angular momentum:
\begin{align}
	x_0 & \equiv \tilde{L}^2 \nonumber,\\
	e_0^2 & \equiv 1 + \tilde{L}^2\left(\frac{E^2}{c^4}-1\right) \,,
	\label{eq:inv}
\end{align}
where we define $\tilde{L}\equiv cL/Gm$ as the dimensionless angular momentum. 
 
 Using the 3PN expressions for $E$ and $L$ given in  equations (A1) and (A2) of \cite{2015PTEP.2015g3E03S} within our Eq.\ (\ref{eq:inv}), we express the variables $e_{\mathrm{BL}}$ and $x_{\mathrm{BL}}$ in terms of $e_0$ and $x_0$ in a PN expansion. These are, in turn, converted to osculating orbit elements $e$ and $x$, utilizing Eqs.\ (3.24) in WM \cite{2017PhRvD..95f4003W} and our Eq.\ (\ref{eq:inv}). 
 
We substitute the orbital element conversion into the Sago-Fujita secular evolution equations to find their equations as a function of osculating orbit elements. 
 
 Next we find the secular evolution equations for $e$ and $x$ with
 \begin{align}
 	\frac{dx}{dt}& = \frac{(dE/dt)(dL/de) - (dE/de)(dLdt)}{(dE/dx)(dL/de) - (dE/de)(dLdx)}\,,
	\nonumber\\
  \frac{de}{dt}& = \frac{(dE/dx)(dL/dt) - (dE/dt)(dLdx)}{(dE/dx)(dL/de) - (dE/de)(dLdx)} \,.
 \end{align}
Finally we convert from $d/dt$ to $d/d\theta$ using $\theta = (2\pi/P)t$ together with the Keplerian expression for the orbital period, arriving at the expressions given in Eqs.\ (\ref{eq:55}).   
It is worth noting that this procedure leads to complete agreement in the $\eta = 0$ limit with SF for the 3.5PN and 4.5PN terms.    The tail terms given in  equations (A1) and (A2) of SF \cite{2015PTEP.2015g3E03S}, are also in agreement with those obtained in  Appendix \ref{app:tails} to the corresponding order in powers of $e$.


 \end{document}